\documentclass[a4paper,11pt]{article}
\usepackage{pos}

\title{Study of the central exclusive production of $\pi^+\pi^-$, $K^+K^-$ and $p \bar{p}$ pairs in proton--proton collisions at $\sqrt{s} = 510$ GeV with the STAR detector at RHIC}
\ShortTitle{Study of the central exclusive production in proton--proton collisions at $\sqrt{s} = 510$ GeV}

\renewcommand{\printHeadAuthors}{T. Truhlář}
\author*[1]{Tomáš Truhlář} 
\note{For the STAR Collaboration.}

\affiliation[]{Czech Technical University in Prague,\\
Faculty of Nuclear Sciences
and Physical Engineering,\\
Brehova 78/7, Prague 115 19, Czech Republic}

\emailAdd{Tomas.Truhlar@fjfi.cvut.cz}

\abstract{We report on the measurement of the central exclusive production process $pp \ \rightarrow \ pXp$ in proton--proton collisions at RHIC with the STAR detector at $\sqrt{\mathrm{s}} = 510$ GeV. At this energy, this process is dominated by a double Pomeron exchange mechanism. The tracks of the centrally produced system $X$ were reconstructed in the central detector of STAR, the Time Projection Chamber and the Time of Flight systems, and were identified using the ionization energy loss and the time of flight method. The diffractively scattered protons, moving intact inside the RHIC beam pipe after the collision, were measured in the Roman Pots system allowing full control of the interaction's kinematics and verification of its exclusivity. The preliminary results on the invariant mass distributions of centrally produced $\pi^+ \pi^-$, $K^+ K^-$ and $p \bar{p}$ pairs measured within the STAR acceptance are presented.
}

\FullConference{%
  40th International Conference on High Energy physics - ICHEP2020\\
  July 28 - August 6, 2020\\
  Prague, Czech Republic (virtual meeting)
}

\begin{document}
\maketitle


\section{Central exclusive production}
Measurement of central exclusive production (CEP)~\cite{CEP} in proton--proton collisions is an important test of quantum chromodynamics. The data from the STAR experiment at RHIC~\cite{RHIC} gives an unique opportunity to perform such studies as it was confirmed by the most recent results of the CEP in proton-proton collisions at $\sqrt{s} = 200$ GeV with the STAR detector~\cite{Rafal}.

\section{Data sample and event selection}
In 2017, the STAR experiment collected proton--proton collision data at $\sqrt{s}= 510$~GeV.
About $622$ million events that were collected using a special CEP trigger system were analyzed.

In order to select a sample of CEP events the following selection criteria were used. First, information from Roman Pot stations~\cite{RP} were checked to ensure that only events with one proton on each side of the interaction point were selected. To ensure good quality of the proton track, all eight silicon planes were required to be used in the proton reconstruction. In addition, the reconstructed proton was required to have transverse momenta inside a fiducial region, as listed in the legend of Fig.~\ref{pipikk} and \ref{pppi}, to ensure high geometrical acceptance. 

Second, only events with exactly two opposite-sign Time Projection Chamber (TPC) tracks matched with two Time of Flight hits and originating from the same vertex were selected. Furthermore, cuts to ensure high geometrical acceptance for the central tracks in the entire fiducial phase space were applied: a cut on the $z-$position of the vertex ($| z\text{-position of vertex}| < 80 $ cm) and a cut on pseudorapidity of central tracks ($| \eta | < 0.7$). Moreover, tracks reconstructed in the TPC had to satisfy track quality cuts -- number of hits on the track and number of hits used for dE/dx.

Finally, to ensure exclusivity of the event, a cut on $p_{\text{T}}^{miss}$ ($p_{\mathsf{T}}^{miss} < 100$ MeV) was applied, where the $p_{\text{T}}^{miss}$ is an absolute value of sum of the transverse momenta of all measured particles. For CEP processes the $p_{\text{T}}^{miss} = 0$ because of the conservation of momentum. 

As a results we obtained 62077 $\pi^+\pi^-$, 1697 $K^+K^-$ and $125$ $p\bar{p}$ CEP event candidates.

\section{Results and summary}
    
\begin{figure}[htbp!]
        \centering
        \includegraphics[width = .45\linewidth]{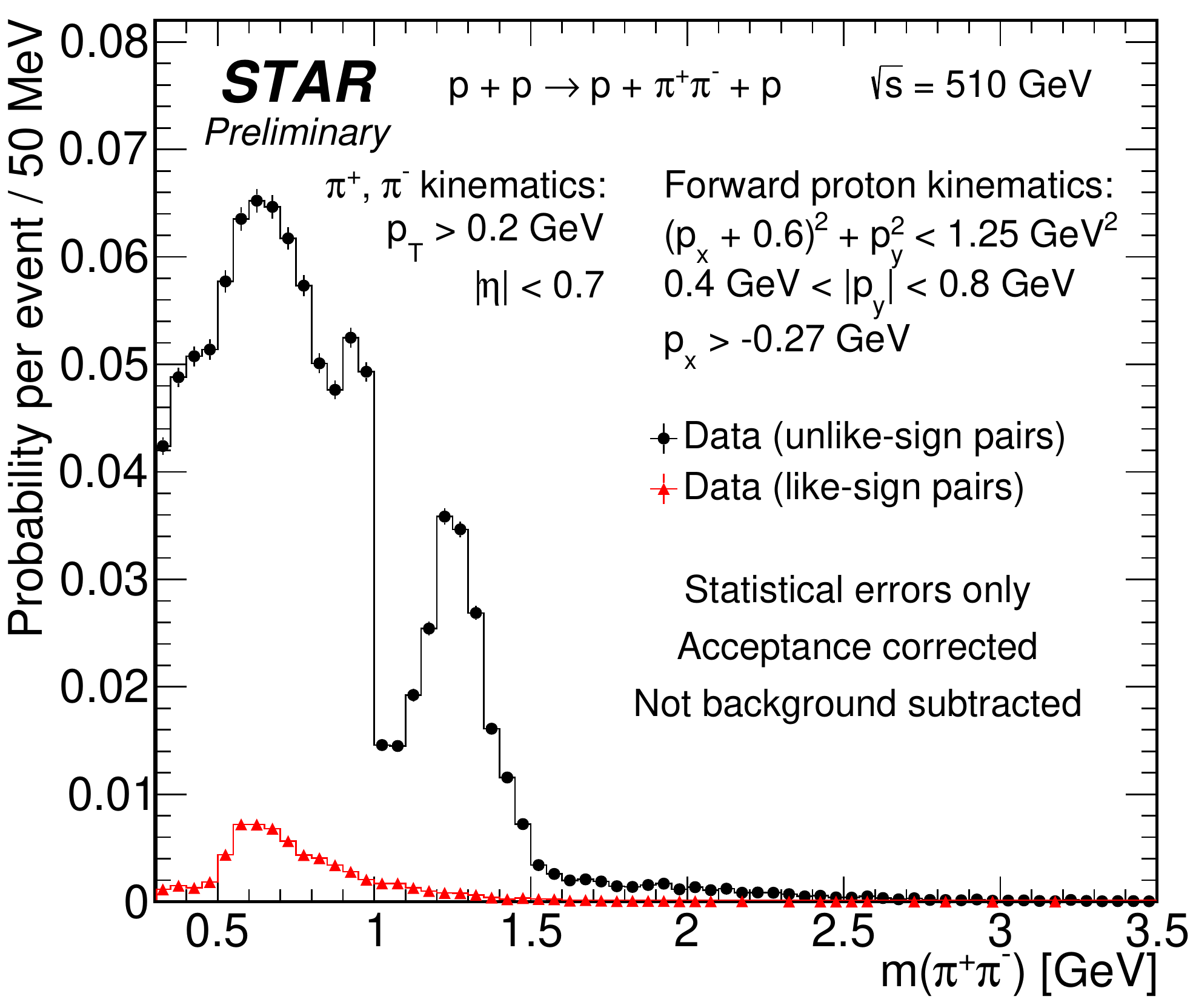}
        \hfill
        \includegraphics[width = .45\linewidth]{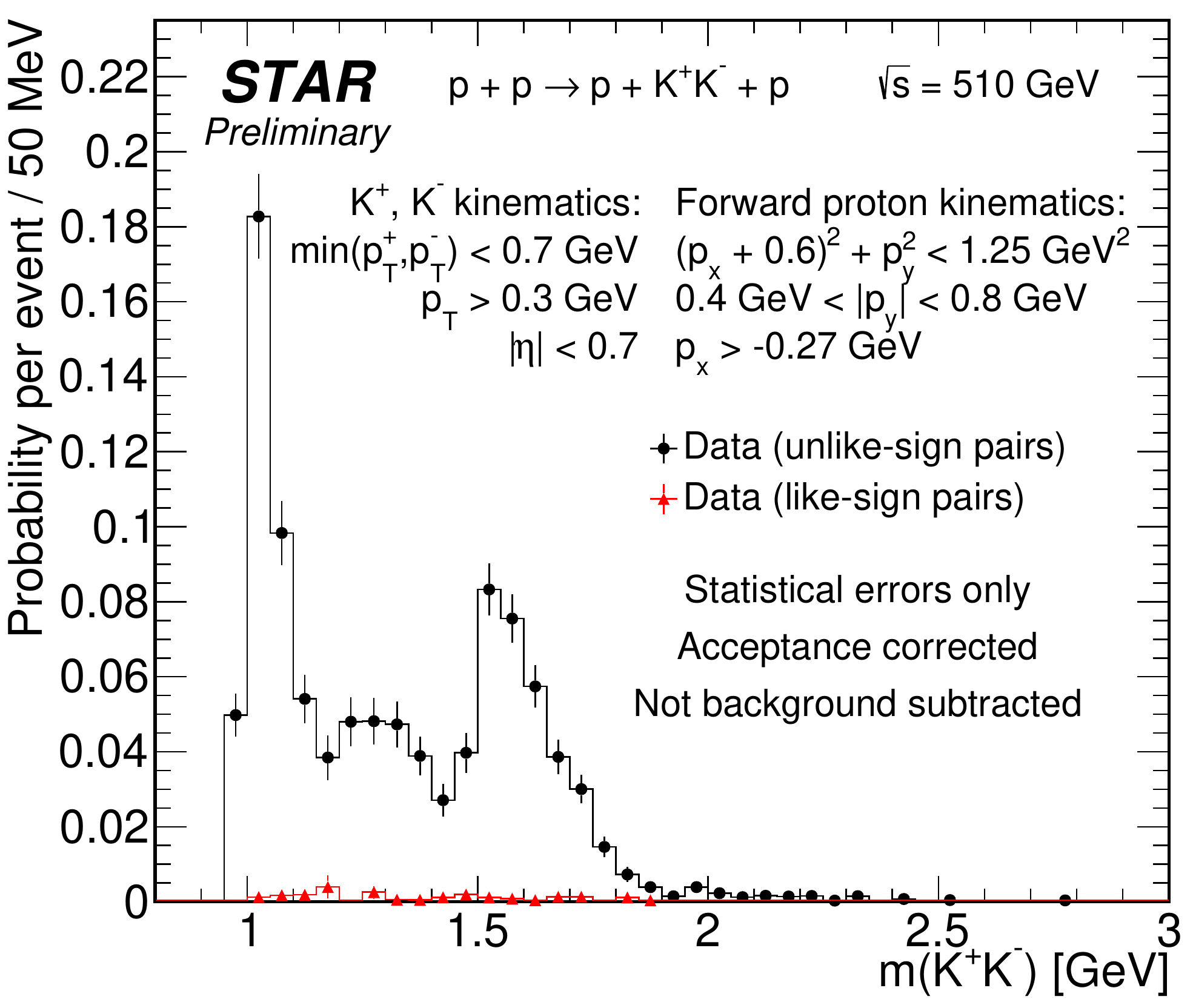}
        \caption[]{The acceptance corrected invariant mass spectrum of exclusively produced $\pi^+\pi^-$ pairs (left) and $K^+K^-$ pairs (right). Error bars represent the statistical uncertainties.}
        \label{pipikk}
    \end{figure}

\begin{figure}[htbp!]
        \centering
        \includegraphics[width = .45\linewidth]{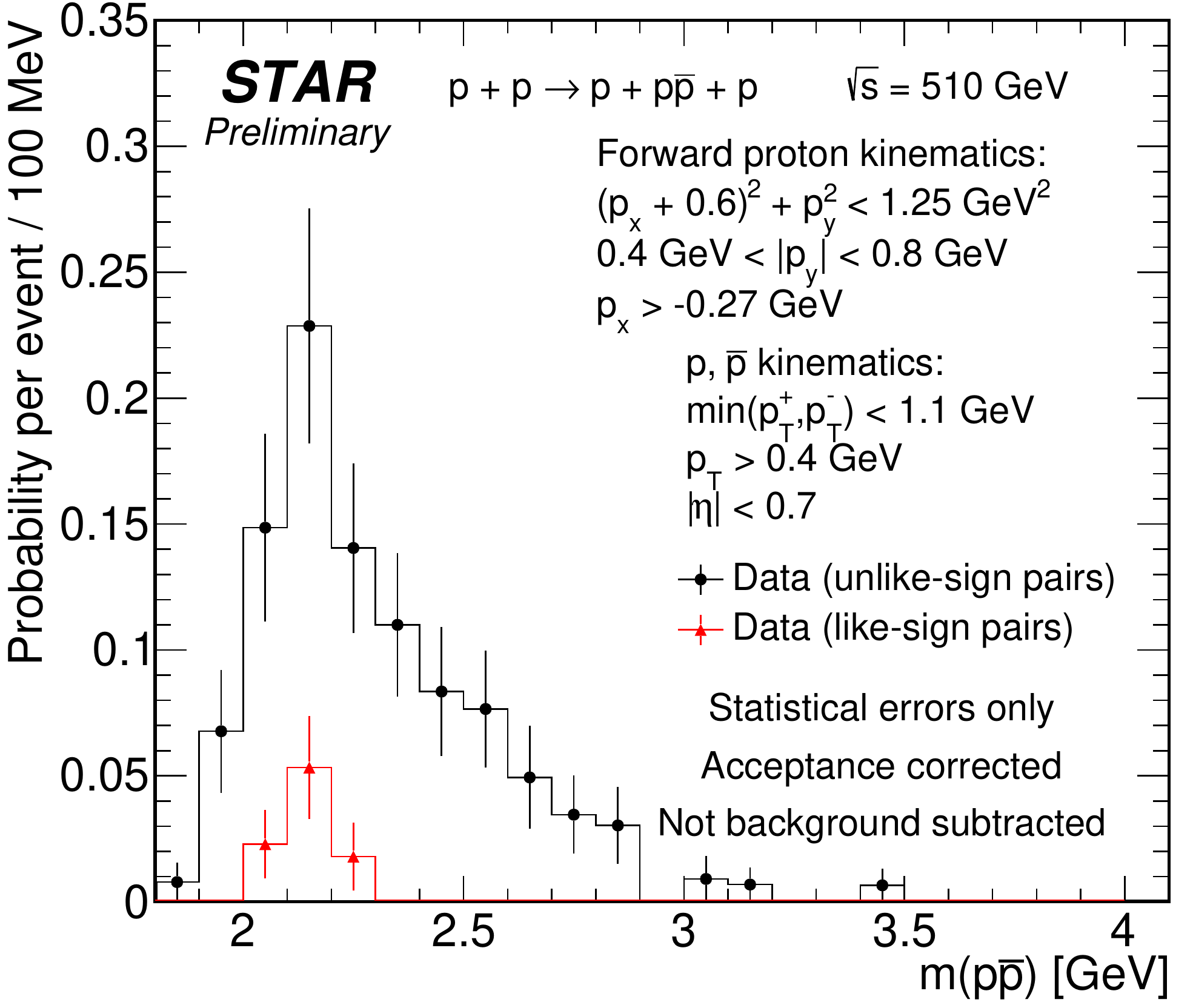}
        \hfill
        \includegraphics[width = .45\linewidth]{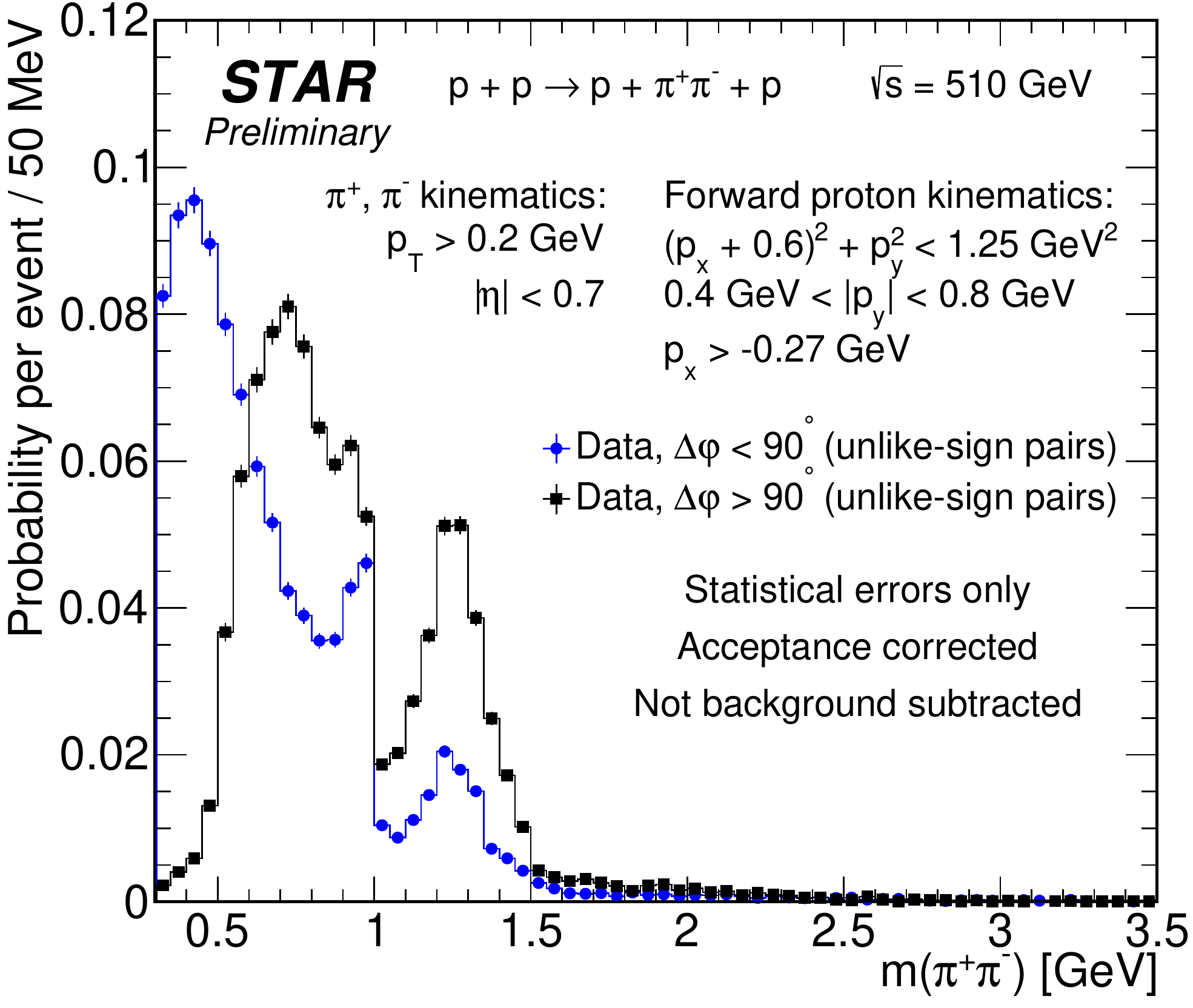}
        \caption[]{The acceptance corrected invariant mass spectrum of exclusively produced $p \bar{p}$ pairs (left). Acceptance corrected invariant mass spectra of exclusively produced $\pi^+\pi^-$ pairs
        in two regions of the difference of azimuthal angles of the forward protons: $\Delta \varphi < 90^{\circ}$ and $\Delta \varphi > 90^{\circ}$ (right). Error bars represent the statistical uncertainties.}
        \label{pppi}
    \end{figure}

    In Figs.~\ref{pipikk} and \ref{pppi}, we present invariant mass distributions of selected $\pi^+ \pi^-$, $K^+ K^-$ and $p \bar{p}$ pairs measured within the STAR acceptance. The results were corrected for acceptance. Fig.~\ref{pppi} (right) shows the invariant mass distribution of $\pi^+\pi^-$ differentiated in two regions of $\Delta \varphi$, where $\Delta \varphi$ is the difference of azimuthal angles of the forward protons. The invariant mass of $\pi^+\pi^-$ shows the expected features, a drop at about 1 GeV and a peak consistent with the $f_2$(1270).

The first results on CEP of $\pi^+ \pi^-$, $K^+ K^-$ and $p \bar{p}$ pairs in proton-proton collisions at $\sqrt{s} = 510$ GeV measured by the STAR experiment have been presented. 
There are ongoing studies of $\pi^+ \pi^-$, $K^+ K^-$, $p \bar{p}$ and $\pi^+ \pi^- \pi^+ \pi^-$ channels with a planned analysis involving the partial wave analysis.

\acknowledgments
This work was supported from the project LTT18002 of the Ministry of Education, Youth, and Sport of the Czech Republic and from European Regional Development Fund-Project "Center of Advanced Applied Science" No. CZ.02.1.01/0.0/0.0/16-019/0000778.

\end{document}